\documentclass[sigconf,screen]{acmart}

\usepackage{subcaption}

\renewcommand\footnotetextcopyrightpermission[1]{} 

\AtBeginDocument{}

\acmConference[OSCAR Workshop]{The 5th Workshop on Open-Source Computer Architecture Research}{June 28, 2026}{Raleigh, USA}

\setcitestyle{acmnumeric}
\newcommand{\toolname}{ORRAM} 
\newcommand\blfootnote[1]{%
  \begingroup
  \renewcommand\thefootnote{}\footnote{#1}%
  \addtocounter{footnote}{-1}%
  \endgroup
}

\begin{document}

\title{\toolname{}: An OpenROAD-Integrated RAM Generator Using Standard Cells }

\author{Brayden Louie*$^1$, Thinh P. Nguyen*$^1$, Matt Liberty$^2$, Austin Rovinski$^1$}
\affiliation{
	\institution{$^1$New York University, $^2$Precision Innovations, Inc.}
}
\email{{brayden.louie,pn2409,rovinski}@nyu.edu, mliberty@precisioninno.com}

\maketitle

\section{Introduction}


Memory inference remains a significant challenge in turnkey ASIC design flows. Inferring flip-flops from RTL can create thousands of densely interconnected instances which dramatically slow down design flows and impede performance. Memory compilers address this issue, although they are third-party tools which are often PDK-specific and may require specialized cells not in the base PDK.

To address these shortcomings, we present \toolname{}, a standard-cell-based memory generator built as a native module within OpenROAD. Given a standard cell library, \toolname{} produces a fully placed and routed RAM block requiring no custom bitcells or external tooling, with timing verification via OpenSTA rather than SPICE simulation. \toolname{} supports arbitrary word sizes, word counts, mask granularities, multi-port read configurations, column muxing, latch-based storage, and automatic PDK-agnostic cell selection, making it compatible with most standard cell libraries including sky130hd and NanGate45. Evaluated on SkyWater 130nm, \toolname{} matches the bit density of historical DFFRAM results while offering a significantly expanded feature set. The source code is available as part of the OpenROAD project~\cite{orram}.

\section{Related Work}
Several open-source alternatives exist for generating on-chip memories, however they each have shortcomings for designers seeking high productivity with low barrier to entry.
The primary option for digital designers is to declare the memory in RTL, which will be inferred as flip-flops during synthesis and then placed and routed with other logic. Open-source tools like Yosys~\cite{yosys} and OpenROAD~\cite{openroad} handle this methodology; however, this technique can incur a high power/performance/area (PPA) penalty due to synthesis algorithms which do not infer buses and generic placement algorithms that reduce logic utilization. Further, runtime of the overall digital flow is impacted by inserting thousands -- possibly millions -- of cells into the netlist and slowing down optimization algorithms.

In contrast, memory compilers can alleviate many of the PPA and runtime issues by generating the memory as a block.
OpenRAM~\cite{OpenRAM} is an open-source SRAM compiler that constructs high-density SRAM arrays; however, it requires designing custom cells like 6T bitcells and sense amplifiers for each PDK being targeted.
Further, it requires expensive SPICE simulations for power and timing info.

DFFRAM~\cite{DFFRAM} instead takes a standard-cell-based approach, constructing RAM arrays from flip-flops/latches and associated logic without requiring custom bitcell design. Previous versions achieved high bit densities through a custom placer. However, DFFRAM now only generates netlists and instead relies on LibreLane~\cite{librelane} for placement and routing, resulting in significantly lower densities.

\toolname{} addresses these limitations by taking a fully integrated approach to standard cell memory generation within OpenROAD. Rather than operating as a standalone compiler, \toolname{}~\cite{orram} generates a fully placed and routed memory block as a native OpenROAD module and performs timing verification via OpenSTA instead of SPICE simulation.

\section{Features and Implementation}
\blfootnote{* denotes equal author effort}
\vspace{-\baselineskip}

\begin{figure}[!t]
    \centering
    \vspace{-2mm}
    \includegraphics[width=\columnwidth, trim=0 250pt 0 100pt, clip]{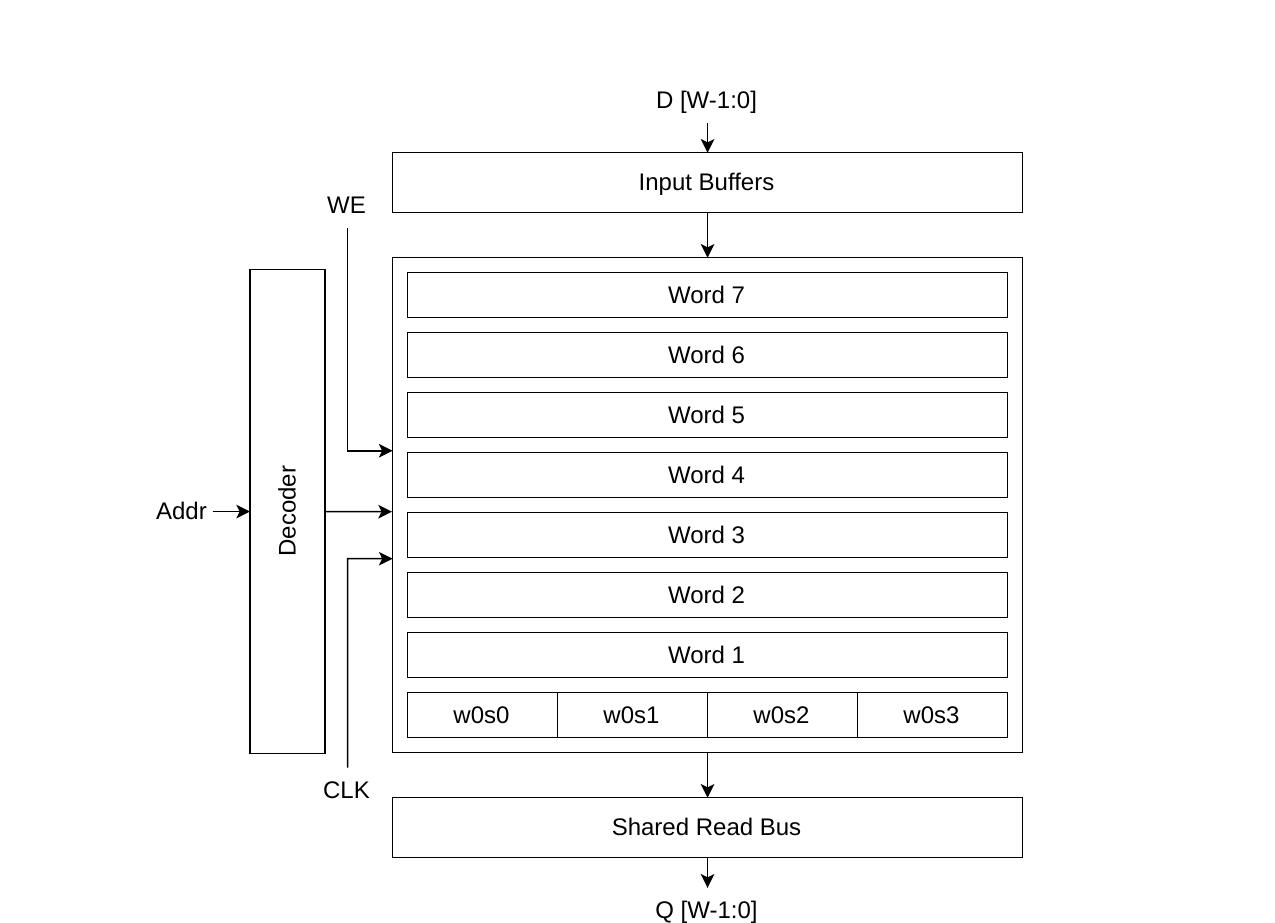}
    \caption{\toolname{} Top-Level Architecture}
    \label{fig:highlevel}
    \vspace{-2mm}
\end{figure}

\begin{figure}[!t]
    \centering
    \includegraphics[width=\columnwidth, trim=40pt 400pt 0 200pt, clip]{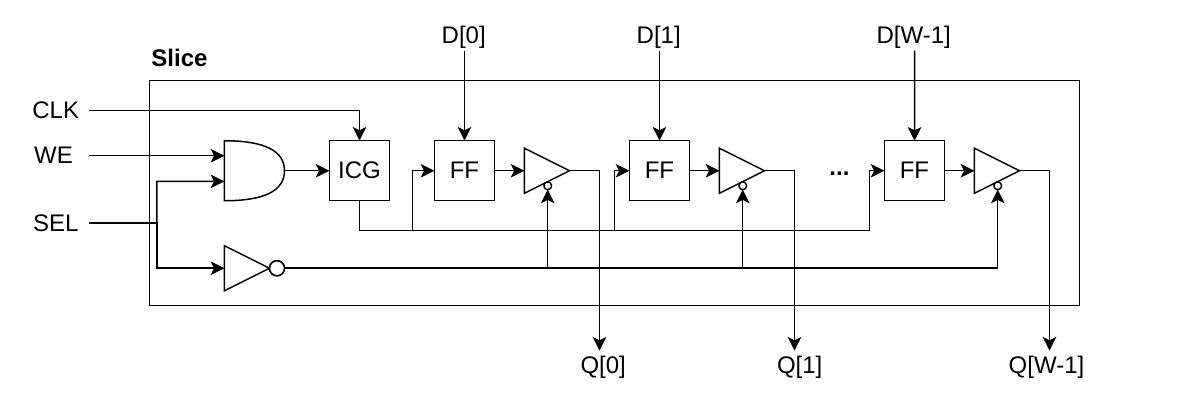}
    \caption{\toolname{} Word Slice Architecture}
    \label{fig:slice}
\end{figure}

\textbf{RAM Structure.} \toolname{}'s default memory architecture consists of an input buffer stage, a word decoder, and a storage array (Fig.~\ref{fig:highlevel}). Input buffers drive data signals into the high-fanout core array, while word selection is handled by a decoder. One decoder is generated per RAM port, generating per-word write enable and per-port read select signals. The array is composed of flip-flops/latches, with tristate cells gating each word's output onto a shared read bus.

\begin{figure}[!t]
    \centering
    \includegraphics[width=\columnwidth]{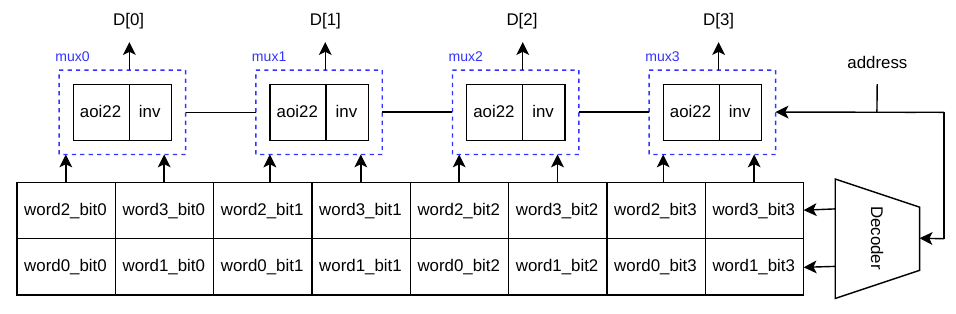}
    \caption{4$\times$4 RAM, \texttt{column\_mux\_ratio}=2}
    \label{fig:mux2}
\end{figure}

\textbf{Column muxing}. Column muxing enables changing the aspect ratio of a RAM by placing multiple words per physical row. This is useful for typical memories which have much larger word counts than word sizes. \toolname{} supports a \texttt{column\_mux\_ratio} of \{1, 2, 4\}, which corresponds to the number of words per physical row. For each physical row, words are interleaved such that each bit is adjacent as shown in Fig.~\ref{fig:mux2}. A row of multiplexers is created at the top of the array using AOI22 bit cells, which are generally available in any standard cell library.
The lowest $\log_2(\texttt{column\_mux\_ratio})$ bits of the address select the word within a row, and the remaining upper bits select the physical row.

    

\begin{figure}[!t]
    \centering
    \includegraphics[width=0.6\columnwidth]{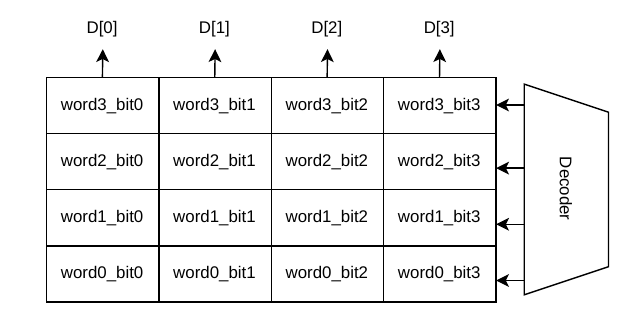}
    \caption{4$\times$4 RAM, \texttt{column\_mux\_ratio}=1}
    \label{fig:mux1}
    \vspace{-2mm}
\end{figure}

\textbf{Latch memory support}. Latch-based memories have considerable advantages over flip-flop-based memories, offering a compact, faster, and more power-efficient alternative. \toolname{} offers the option to use latches as storage cells through a two-phase latch setup. When enabled, a row of negative latches is inserted between the input buffer and the storage rows, which now consist of positive latches. During the low clock phase, the negative latch in each column captures and holds the input data; on the rising edge, the positive latches in the cells below capture this data, thus having identical functionality to a positive flip-flop. With column muxing enabled, the negative latch position within each bit group is adjusted to maintain balanced fanout across all storage cells below.

\textbf{Behavioral Model.} \toolname{} supports generatiion of a cycle-accurate, PDK-agnostic behavioral model for all configurations. This behavioral model can be used in RTL simulations for significantly faster simulation than simulating directly with the netlist.

\begin{figure}[h]
    \centering
    \includegraphics[width=0.9\columnwidth]{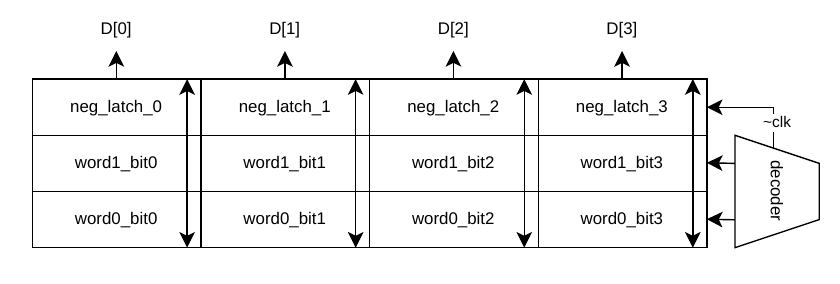}
    \caption{4$\times$4 latch-based RAM Layout}
    \label{fig:latch-layout}
\end{figure}

\section{Results and Takeaways}

\begin{table}[!t]
\caption{Bit density (bits/mm$^2$) across memory generators on SkyWater 130nm.$^{1}$}
\label{tab:comparison}
\footnotesize
\setlength{\tabcolsep}{5pt}
\begin{tabular}{l|c|c|c|c|c|c}
\toprule
\textbf{Size} & \textbf{ORRAM} & \textbf{ORRAM} & \textbf{DFFRAM} & \textbf{DFFRAM} & \textbf{Open-} & \textbf{RTL+} \\
 & \textbf{Latch}$^{2}$ & \textbf{DFF}$^{2}$ & \textbf{(historical)}$^{3}$ & \textbf{(current)} & \textbf{RAM}$^{4}$ & \textbf{ORFS}$^{5}$ \\
\midrule
512\,B & 31,795 & 27,734 & 26,557 & 14,691 & 38,299 & 15,196 \\
1\,KB  & 31,552 & 28,168 & 26,027 & 14,951 & 77,499 & 15,250 \\
2\,KB  & 31,977 & 28,339 & 26,297 & 14,817 & 105,767 & 15,639 \\
4\,KB  & 32,284 & 28,586 & 26,196 & 14,948 & 120,240 & 15,566 \\
8\,KB  & 32,508 & 28,677 & 26,229 & 14,879 & 148,413 & 15,649 \\
\bottomrule
\end{tabular}
\vspace{3pt}
\begin{minipage}{\linewidth}
\footnotesize
$^{1}$ Configured with 64-bit words and byte-level masking.\\
$^{2}$ \toolname{} results use configurations yielding the highest bit density\\
$^{3}$ Historical results reported from the DFFRAM repository; exact configurations are unverifiable as the custom placer has since been removed.\\
$^{4}$ Dimensions reported are raw compiler output, OpenRAM macros are typically wrapped before being used in automated PnR flows.\\
$^{5}$ Configured for 60\% core utilization.
\end{minipage}
\end{table}

We evaluated \toolname{} on sky130hd across a range of memory sizes from 512 bytes to 8 KB, compared to other current open-source alternatives. All configurations target 64-bit word reads with byte-level write masking (Table~\ref{tab:comparison}).

\toolname{}'s DFF-based memory achieves bit densities of approximately 28,000 bits/mm$^2$ across all tested sizes, representing roughly 2× the density of the current DFFRAM project, which achieves approximately 14,000–15,000 bits/mm$^2$ across the same configurations. This gap stems from DFFRAM relying on LibreLane for placement and routing using generic algorithms, whereas \toolname{} uses OpenROAD's API to create a custom placer and enable much higher placement regularity and density. Notably, historical DFFRAM results using its deprecated custom placer achieved approximately 26,000–26,500 bits/mm$^2$, which \toolname{}'s DFF-based approach matches while offering a significantly expanded feature set. By comparison, the behavioral RTL approach that uses OpenROAD Flow Scripts (ORFS) achieves approximately 15,000–15,600 bits/mm$^2$, and additionally incurs the overhead of executing the full RTL-to-GDS flow including synthesis, clock tree synthesis, and general-purpose placement and routing. OpenRAM achieves higher peak densities at larger sizes (up to $\sim$148,000 bits/mm$^2$ at 8 KB) due to its custom 6T bitcell layouts, but comes with the need for hand-designed cells for each target PDK and cumbersome SPICE simulations to measure timing.

\toolname{}'s latch-based memory provides a 15\% density improvement over its DFF counterpart, achieving 31,000–32,000 bits/mm$^2$ across tested sizes due to the smaller footprint of latch cells relative to flip-flops, with the trade-off of additional timing analysis complexity due to the two-phase capture behavior. For designs where latch-based timing closure is acceptable, this configuration offers the highest density of any standard-cell-based approach evaluated.

Beyond density, \toolname{} provides several capabilities absent from the current DFFRAM project: multi-port read/write configurations,
configurable column muxing for aspect ratio control, and native integration within OpenROAD and OpenROAD Flow Scripts. Combined, these features make \toolname{} a drop-in memory generation solution for open-source ASIC design without requiring custom bitcell libraries or an external flow setup.

\section{Future Work}
While \toolname{} already offers many features to support faster design space exploration and even tapeouts, we are still working on several features to improve usability and performance including: 1) tightly integrating OpenSTA to perform timing and power-based optimization on RAM configurations, 2) banking support to help support even larger-scale RAMs, and 3) automatically generating RAMs from the memory inference step during synthesis.

\bibliographystyle{ACM-Reference-Format}
\bibliography{ref}

@misc{yosys,
  author       = {Claire Wolf},
  title        = {{Yosys} Open {SYnthesis} Suite},
  howpublished = {\url{https://yosyshq.net/yosys/}}
}

@inproceedings{openroad,
  title={{OpenROAD: Toward a Self-Driving, Open-Source Digital Layout Implementation Tool Chain}},
  author={Ajayi, Tutu and Blaauw, David and Chan, Tuck-Boon and Cheng, Chung-Kuan and Chhabria, Vidya A. and Choo, David K. and Coltella, Matteo and Dobre, Sorin and Dreslinski, Ronald G. and Fogaça, Mateus and Hashemi, Soheil and Hosny, Abdelrahman and Kahng, Andrew B. and Kim, Minsoo and Li, Jiajia and Liang, Zhaoxin and Mallappa, Uday and Penzes, Paul and Pradipta, Geraldo and Reda, Sherief and Rovinski, Austin and Samadi, Kambiz and Sapatnekar, Sachin S. and Saul, Lawrence and Sechen, Carl and Srinivas, Vaishnav and Swartz, William and Sylvester, Dennis and Urquhart, David and Wang, Lutong and Woo, Mingyu and Xu, Bangqi},
  booktitle = {Proc. GOMATECH},
  year = {2019}
}

@misc{orram,
  title = {{ORRAM}},
  author = {Louie, Brayden and Nguyen, Thinh P. and Liberty, Matt and Rovinski, Austin},
  url = {https://github.com/The-OpenROAD-Project/OpenROAD/tree/master/src/ram},
  year = {2026}
}

@inproceedings{OpenRAM,
  title     = {{OpenRAM}: An Open-Source Memory Compiler},
  author    = {Guthaus, Matthew R. and Stine, James E. and Ataei, Samira and Chen, Brian and Wu, Bin and Sarwar, Mehmet},
  booktitle = {Proceedings of the IEEE/ACM International Conference on Computer-Aided Design (ICCAD)},
  year      = {2016},
  doi       = {10.1145/2966986.2980098}
}

@misc{DFFRAM,
  title        = {{DFFRAM}: Standard-Cell-Based Memory Compiler},
  author       = {Shalan, Mohamed and Gaber, Mohamed and Abdelatty, Manar and Nofal, Ahmed and Moussa, Hany and Moussa, Ramez and {Efabless Corporation}},
  year         = {2021},
  howpublished = {\url{https://github.com/AUCOHL/DFFRAM}}
}

@inproceedings{librelane,
  author    = {Shalan, Mohamed and Edwards, Tim},
  booktitle = {2020 IEEE/ACM International Conference On Computer Aided Design (ICCAD)},
  title     = {Building {OpenLANE}: A 130nm {OpenROAD}-based Tapeout-Proven Flow: Invited Paper},
  year      = {2020},
  pages     = {1-6}
}

\end{document}